\documentclass{vgtc}                          

\graphicspath{{figures/}{pictures/}{images/}{./}} 

\usepackage{times}                     

\usepackage{tabu}                      
\usepackage{booktabs}                  
\usepackage{lipsum}                    
\usepackage{mwe}                       

\usepackage{mathptmx}                  

\usepackage{amsmath}
\usepackage{amssymb}

\usepackage{mathptmx}                  
\usepackage{microtype} 
\onlineid{0}

\vgtccategory{Research}

\vgtcinsertpkg

\usepackage{hyperref} 
\usepackage[most]{tcolorbox}

\definecolor{review}{rgb}{0, 0, 0}

\title{CATVis: A Collaborative Multi-Agent Workflow for Turbomachinery Simulation Data Visualization}

\author{
  Zhe Wang$^{1,2*}$,
  Zehao Lou$^{2,3}$\thanks{Authors contributed equally.},
  Guanghui Zhao$^{2,3}$,
  Yu Dong$^{2}$,
  Guan Li$^{1,2}$,\\
  Pengyi Xu$^{4}$,
  Gaorong Liang$^{1}$,
  Jun Liu$^{1,2}$,
  Guihua Shan$^{1,2,3}$\thanks{Corresponding author: sgh@cnic.cn}\\[5pt]
  \footnotesize 
  \begin{tabular}{ll}
    $^{1}$ University of Chinese Academy of Sciences, Beijing, China & 
    $^{2}$ Computer Network Information Center, Chinese Academy of Sciences \\
    $^{3}$ Hangzhou Institute for Advanced Study, UCAS, Hangzhou, China & 
    $^{4}$ Michigan Technological University, Houghton, MI, USA
  \end{tabular}
}

\teaser{
  \centering
  \includegraphics[width=\linewidth]{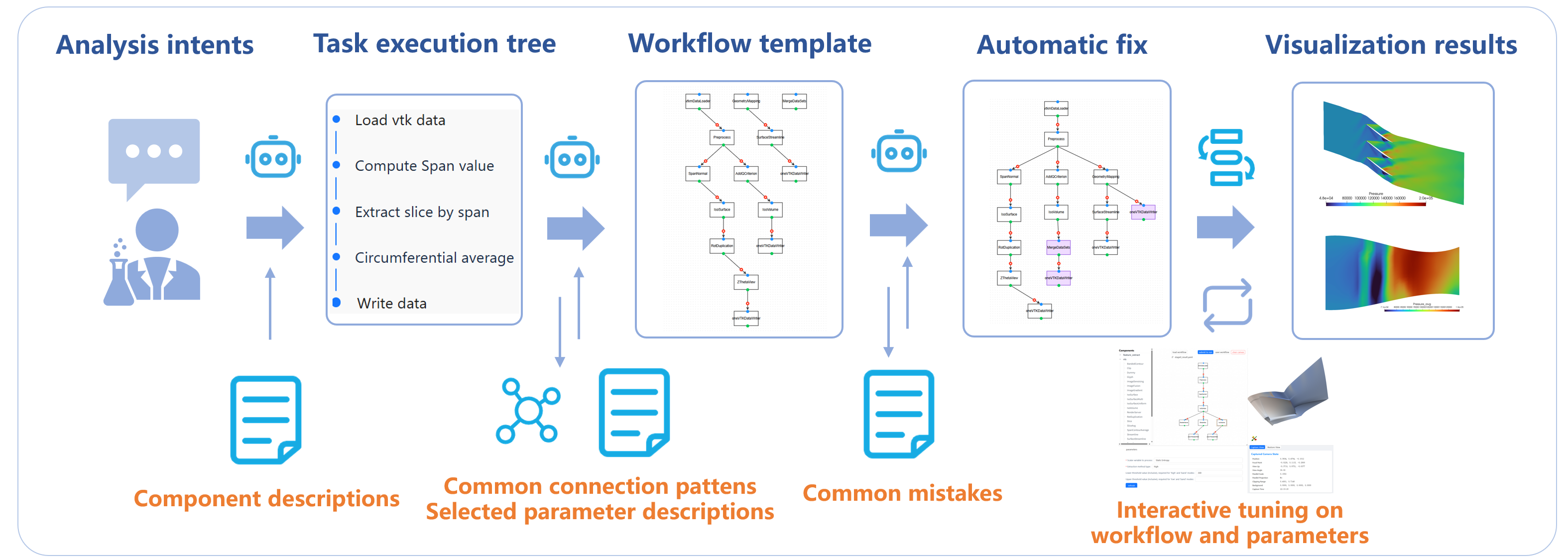}
  \caption{
Overview of CATVis: 
(1) User intents are parsed into a task execution tree composed of domain-specific task components derived from turbomachinery visualization algorithms.
(2) The execution tree is mapped to a structured workflow template represented as a DAG, guided by component descriptions and common connection patterns.
(3) The generated workflow is automatically diagnosed and refined using a knowledge base of common error patterns. 
(4) An interactive interface enables human-in-the-loop refinement, allowing users to adjust workflow structure and parameters (e.g., camera settings).
  }
  \label{fig:teaser}
}

\abstract{

Recent advances in AI for Science have enabled natural language (NL) interfaces for scientific data analysis. In turbomachinery CFD post-processing, translating ambiguous high-level analytical goals (e.g., vortex identification) into precise visualization procedures supporting complex domain-specific analysis is challenging.
We present CATVis, \textcolor{review}{a} \textbf{C}ollaborative multi-\textbf{a}gen\textbf{t} workflow system that bridges this gap by transforming NL intents into structured middle representation for visualization. Our approach reformulates domain-specific visualization procedures as composable workflow representations, and use multi agent to generate workflow representations via intent planning, template generation, and error-aware refinement, where each stage incrementally updates a shared structured representation.  
We evaluate the impact of external knowledge and workflow structuring on generation accuracy, demonstrating that the proposed approach significantly improves complex workflow generation correctness while reducing prompt complexity.

} 

\begin{document}


\firstsection{Introduction}

\maketitle

\begin{figure*}[h]
    \centering 
    \includegraphics[width=0.95\textwidth]{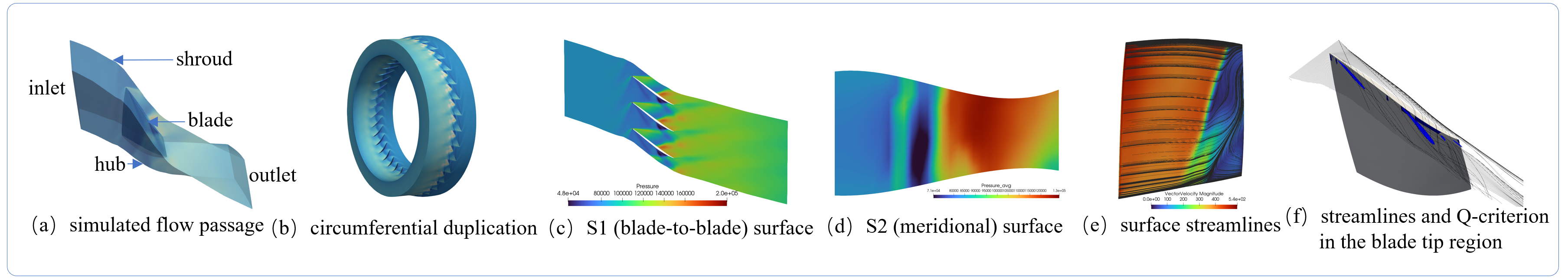}  
    \caption{Visualization results for compressor simulation data, generated by our workflow system and rendered in ParaView.}
    \label{fig:motivation2}
    \vspace{-15pt}
\end{figure*}

AI for Science (AI4Science) has emerged as a new paradigm for accelerating scientific discovery, with LLM-based agents serving as a central component across a growing range of scientific software applications~\cite{Ren2025TowardsSI}. In particular, data agents have demonstrated promising capabilities in automating data analysis and visualization tasks, mapping natural language descriptions to visualization outputs, including scientific charts~\cite{10121440} and 3D visualizations~\cite{zhu2025survey}. The rapid advancement of pretrained LLMs, combined with improved knowledge organization techniques, has further accelerated this trend~\cite{shen2022towards}. Despite this progress, generating complex, domain-specific visualization pipelines remains challenging.

Post-processing of Computational Fluid Dynamics (CFD) simulations is essential for understanding complex flow phenomena across a wide range of engineering applications. CFD solvers produce high-dimensional, unstructured datasets requiring sophisticated visualization pipelines for interpretation. Among these applications, turbomachinery simulation post-processing plays a critical role in validating the design of engine components, including compressors, turbines, and combustion chambers. Unlike general-purpose CFD visualization, turbomachinery post-processing relies on specialized techniques that exploit geometric information intrinsic to the rotating machinery. As examples shown in Figure~\ref{fig:motivation2}, blade-to-blade surfaces at a constant spanwise height and meridional plane through the machine axis require coupling the flow solution with the underlying blade and passage geometry.

Commercial post-processing software, such as CFX-Post~\cite{ansys_turbo_selector_2024} and Tecplot Macro Language~\cite{Tecplot2020}, provides dedicated turbomachinery modules that leverage geometric metadata during visualization. 
They target interactive use by domain experts and provide domain-specific scripting interfaces for visualization filters and operations.
\textcolor{review}{However, low-level macro languages (e.g., Tecplot~\cite{heidbach2020manual}) are inherently state-dependent and structurally rigid; without a high-level abstraction layer, they force LLMs to manage abstract planning and intricate execution syntax simultaneously, increasing the difficulty of reliable generation.}
Open-source visualization platforms such as ParaView~\cite{ayachit2015paraview} provide Python-based scripting interfaces, and recent studies~\cite{peterka2025chatvis,liu2025paraview} have demonstrated the feasibility of integrating LLMs with the open source tool for automated visualization generation. 
\textcolor{review}{However, these studies are developed on mature ParaView and VTK ecosystems, where abundant public resources enable LLMs to leverage existing knowledge learned during pretraining.}
In addition, they are not designed for complex turbomachinery post-processing workflows that involve multiple geometry-aware operations with dependency relationships, such as span-value computation, duplication, and blade-to-blade visualization. 
\textcolor{review}{
Since turbomachinery visualization libraries lack robust public documentation, rapidly changing APIs, and highly specialized operations, ensuring that LLMs accurately invoke these evolving interfaces remains challenging.}





\textcolor{review}{
To address the difficulty of reliably generating workflows over low-level scripting interfaces, we abstract turbomachinery visualization algorithms into composable workflow tasks with structured composition rules. This abstraction reduces generation complexity and enables interactive human refinement. To further improve generation quality and support flexible context organization, we decompose workflow generation into multiple collaborative stages, where stage-specific contextual knowledge supports intent planning, workflow construction, and feedback-driven refinement. The main contributions of this paper are listed as follows:
}

\noindent\textbf{(1) Workflow-centric abstraction of complex turbomachinery visualization processes.}
We abstract domain-specific turbomachinery procedures into composable units organized as directed acyclic graphs (DAGs). Each unit exposes structured attributes that describe its semantics and parameters. We further design an intermediate workflow representation that balances semantic clarity, human editability, and LLM generation efficiency.


\noindent\textbf{(2) A collaborative multi-agent workflow generation framework.}
We propose a multi-agent visualization generation approach that constructs structured visualization workflow templates from user intents. The framework supports human-in-the-loop refinement through an interactive interface, allowing users to efficiently correct and improve generated workflows.


\noindent\textbf{(3) Systematic evaluation of workflow generation and context strategies.}
We evaluate the ability of pretrained LLMs to translate visualization intents into executable workflow (DAG representations) based on published turbomachinery use cases, and analyze how different context organization strategies affect generation accuracy and workflow quality.


\section{Background and related work}

In this section, we first discuss the visualization of turbomachinery simulation data, and then present the multi-agent framework and the research gap for visualizing it.

\subsection{Typical Turbomachinery Visualization}

Visualization is an essential component of CFD post-processing~\cite{ilieva2019cfd}, enabling domain experts to analyze complex flow behaviors through interactive visual exploration. Typical compressor simulations are performed on a single blade passage 
using periodic boundary conditions, with the computational domain defined by the hub, shroud, inlet, and outlet surfaces (Figure~\ref{fig:motivation2}(a)). The full annular configuration is reconstructed by circumferentially replicating the simulated passage about the rotational axis (Figure~\ref{fig:motivation2}(b)). The S1 surface (Figure~\ref{fig:motivation2}(c)) and S2 surface (Figure~\ref{fig:motivation2}(d)) provide sectional views that enable global inspection of flow patterns within the compressor. Surface streamlines (Figure~\ref{fig:motivation2}(e)) are generated by mapping velocity fields onto blade surfaces and computing constrained streamline trajectories, enabling detailed analysis of flow evolution along blade geometry. In the blade tip region, vortex structures are identified by the Q-criterion combined with isosurface extraction (Figure~\ref{fig:motivation2}(f)), revealing the spatial distribution of tip leakage vortices.

\subsection{Research Trends in Turbomachinery Visualization}
Turbomachinery visualization has been extensively studied over the past decades, particularly in flow visualization and vortex detection~\cite{roth1996flow}. These advances have been incorporated into commercial software systems such as CFX-Post~\cite{ansys_turbo_selector_2024} and Tecplot~\cite{Tecplot2020}, which provide specialized modules for domain-specific visualization, including spanwise slicing, meridional surface extraction, etc.

With the increasing scale of CFD simulations~\cite{10485184}, research works focused on large-scale and in-situ visualization techniques to reduce data transfer overhead and improve efficiency of analysis~\cite{list2007high,7539561} and high fidelity rendering~\cite{Prabhakar2022VirtualCO}. These approaches significantly reduce storage requirements and improve scalability for large-scale turbomachinery simulations. As simulation complexity continues to grow, visualization pipelines have evolved into multi-stage workflows that require repeated configuration of domain-specific operations. Traditional scripting-based approaches partially automate these workflows~\cite{hossain2020vizsciflow,naha2025abstract}, but still rely heavily on expert knowledge to design correct processing sequences.

Recently, CFD research has begun exploring AI-driven workflow automation. Large language models (LLMs) now assist in simulation setup, execution, analysis, and post-processing, as demonstrated by OpenFOAMGPT~\cite{pandey2025openfoamgpt}, ChatCFD~\cite{chatcfd}, and MetaOpenFoam~\cite{chen2025metaopenfoam}. Since visualization-based post-processing is essential in CFD pipelines, these advances suggest that intelligent workflow generation can enhance 
turbomachinery visualization.

\subsection{Agentic workflow for scientific visualization}

Recent research has demonstrated the ability to generate visualization programs using LLMs~\cite{peterka2025chatvis,biswas2025vizgenie,zhao2026toward}. 
These works leverage techniques such as retrieval-augmented generation (RAG), fine-tuning, and in-context learning to improve the reliability and efficiency of visualization code generation. 
Domain-specific systems, such as AuraGenome~\cite{zhang2025auragenome}, further show how LLMs can facilitate visualization workflows tailored to specialized scientific fields. While LLM-based approaches show strong potential for visualization 
generation, complex workflows often require coordinated multi-step 
reasoning, motivating recent exploration of multi-agent approaches for scientific visualization. For example, LightVA~\cite{10753451} demonstrates how LLM-based agents can decompose high-level analysis goals into executable tasks through collaborative planning and execution. Recent studies~\cite{lu2026agentic} extend this idea to multi-agent collaboration for visualization generation and result explanation.

More broadly, agentic workflow typically organizes agents into components such as planners, memory modules, action spaces, and verification mechanisms~\cite{Ren2025TowardsSI}. Recent studies further highlight the importance of suitable intermediate workflow representations and human-in-the-loop mechanisms, such as interactive workflow graphs~\cite{wang2025aop} and human-guided refinement strategies~\cite{tang2026vividoc}.

In summary, despite recent progress in LLM-driven visualization and agentic workflows, \textcolor{review}{they mainly focus on improving interaction mechanisms without explicitly bridging domain-specific CFD simulation data visualization queries and scientific visualization workflow generation. Reliably generating such complex workflows via an Agent-based approach for turbomachinery CFD post-processing remains a major challenge.} 
In particular, there is a lack of structured intermediate representations that facilitate both reliable automated generation and human-in-the-loop interactive refinement.

\section{Method}

This section presents the design of the proposed workflow generation framework. 
We first describe the workflow representation used to organize visualization operations. 
Next, we describe how we prepare context knowledge to support LLM reasoning.
Finally, we present the multi-stage workflow for template generation.

\subsection{Visualization Workflow Representation}

Our visualization workflow is represented as a directed acyclic graph (DAG), where nodes represent visualization filters and edges denote data dependencies between operations. 
Each node encapsulates a specific visualization or data-processing operation, and its parameters are organized using a standardized schema (e.g., PyDantic), which can be serialized to structured formats such as YAML or JSON. This representation bridges domain-specific visualization procedures and LLM-based generation.

The workflow composer parses workflow templates into executable DAGs and schedules node execution according to dependency relationships. 
Domain-specific filters for turbomachinery visualization are integrated into the workflow, including span computation based on hub and shroud geometry, vortex-related scalar computation (e.g., Q-criterion), and streamline generation. Figure~\ref{fig:motivation2} illustrates representative visualization results using the presented workflow system. In addition to domain-specific operations, the workflow incorporates commonly used visualization filters such as contouring, slicing, and offscreen rendering. 
The current library contains more than 50 filters. 
We extend the visualization algorithms introduced in GPVis~\cite{shan2019gpvis} 
and reorganize them into a modular workflow-compatible library.

All workflow tasks inherit from a unified parent task class, enabling consistent parameter management and automatic data connection. 
\textcolor{review}{The graph-based structure increases flexibility for humans to verify the correctness of dependencies in the LLM-generated workflow tasks,} while the structured parameter representation makes the workflow interpretable by both machines and users. 
Users can further modify parameters through the graphical interface when needed. A detailed workflow example and its visualization results are provided in the supplementary materials.


\subsection{Contextual Construction and Prompt Generation}

To support reliable workflow generation, we prepare multiple types of domain-specific knowledge as contextual input to the LLM. 
This contextual information helps the LLM understand workflow construction rules and data details and generate executable visualization pipelines from user queries.

We organize the context knowledge into five categories: (1) workflow examples: representative workflow instances that 
demonstrate common visualization procedures; (2) workflow rules: structural constraints defining task dependencies, data connections, field usage, and valid operator sequences; (3) task descriptions: semantic descriptions of each visualization operation, including its purpose and applicable scenarios; (4) parameter specifications: detailed definitions of task 
parameters, automatically extracted from Python docstrings and PyDantic schemas, and (5) dataset descriptions: metadata describing dataset structure, including field names, block organization, and geometric components (e.g., blade, hub, and shroud regions). This structured contextual information enables the LLM to reason effectively about workflow parameters.
\textcolor{review}{Section 2 of the supplementary materials further illustrates the strategy for preparing contextual information to support multi-agent tasks with different objectives.
}

\subsection{Multi-stage Workflow template generation}
\label{sec:multistages}

We adopt \textcolor{review}{a} multi-stage framework in which multiple agents collaboratively construct executable visualization workflows through progressive refinement (described in Figure~\ref{fig:teaser}). 
Each stage is modeled as an independent agent with a specialized role, including intent interpretation, workflow construction, and error-aware refinement. Agents shared a structured workflow representation that is incrementally updated to ensure consistency and correctness.

The intent agent first parses the user’s natural language input into high-level visualization tasks. 
These tasks are expanded into a task execution tree based on predefined task components, where candidate visualization units are identified to guide downstream generation. 
The workflow generation agent then maps the execution tree into structured workflow templates by instantiating operators and establishing data dependencies. 
To ensure completeness, predefined connection patterns derived from prior examples are applied to insert necessary intermediate operations when dependencies are missing. \textcolor{review}{The refinement agent further improves the generated workflows through error-aware feedback based on predefined function calls to provide suggestions, including common failure patterns collected during development. Section 2 of the supplementary materials provides details of context preparation for each agent.}

Finally, users can interactively inspect and adjust the generated workflows through the interface. 
Parameters that are difficult to specify in natural language (e.g., camera settings) can be refined with visual feedback, and users may modify workflow topology by editing nodes and their associated parameters.

\section{Evaluation}

We carefully select five representative visualization use cases derived from published turbomachinery studies. \textcolor{review}{Table 2 in the supplementary materials provided detailed descriptions for these tasks.}
For each case, we analyze the visualization results reported in the corresponding literature, and reconstruct equivalent reference workflows using our system and simulation data based on the NASA Rotor37~\cite{denton1997lessons}.
We conduct two experiments with these use cases: one evaluating the effectiveness of external knowledge, and the other evaluating the proposed multi-agent workflow for visualization generation. To ensure consistent evaluation, we use DeepSeek-V3 as the base LLM.
\subsection{External Knowledge Ablation}
\label{sec:exp1}

\begin{figure*}[htp]
    \centering 
    \includegraphics[width=0.95\textwidth]{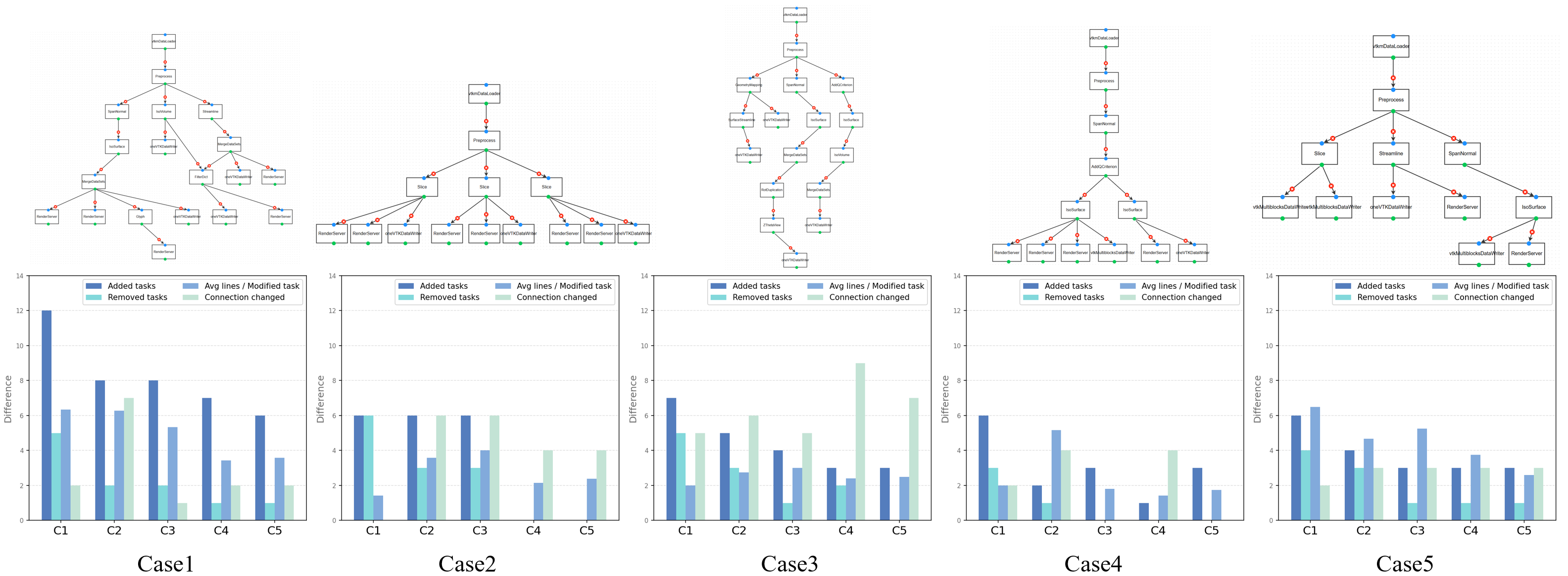}  
    \caption{Comparison of workflow generation errors across five turbomachinery cases under different external knowledge configurations.}
    \label{fig:exp1_diff}
    \vspace{-10pt}
\end{figure*}








This experiment aims to investigate how different combinations of external knowledge affect the quality of workflow template generation from natural language descriptions. \textcolor{review}{Table 1 in the supplementary materials} shows five configurations with increasing levels of external information included in system prompt. \textbf{C1} provides only simple workflow examples (e.g., small pipelines with 3-5 nodes), serving as the baseline to assess how well the model can generalize from minimal structural demonstrations without additional semantic. \textbf{C2} additionally introduces construction rules that define how operators compose valid workflows, giving the model explicit structural constraints.
\textbf{C3} further adds semantic descriptions for each operator, enabling the model to better understand their capabilities.
\textbf{C4} further adds parameter-level descriptions that specify the name and type of arguments accepted by each operator.
\textbf{C5} additionally adds data-level descriptions, including geometric properties and available field variables, ensuring generated workflow parameter consistency with the input data.


We quantitatively compare generated workflows of C1-C5 with reference workflows using \textbf{four metrics}: the number of added tasks, deleted tasks, modified connections, and parameter mismatches. Figure~\ref{fig:exp1_diff} shows the performance of the LLM across five complex turbomachinery visualization cases under five external knowledge configurations, where lower values indicate closer agreement with the ground truth (user prompt of each case are listed in supplementary materials). Based on quantitative trends and qualitative log analysis, we summarize the following key observations:

Across all cases, adding structural rules and operator descriptions 
gradual reduces added and removed tasks from C1 to C5. 
In several cases (e.g., Case2 and Case4), structural errors are fully eliminated under C4 and C5, indicating that the generated workflows become increasingly consistent with the ground truth topology. 
Despite the improvements observed in C5, parameter differences tend to plateau rather than reaching zero. The remaining discrepancies mainly occur in rendering-related parameters, such as camera position, focal point, and view direction, which are inherently difficult to maintain consistently in one-shot generation.
We also observe that performance improvements come at the cost of significantly larger context inputs, including irrelevant operator input parameter descriptions. In practice, when we update the configuration associated with a specific operator, we only need to provide information associated with this operator instead of the whole context.
These observations motivate a design choice that decomposes context information into multiple stages aligned with different generation goals, as evaluated in the next experiment.

\subsection{Multi-Agent Workflow Generation}

This experiment evaluates the proposed multi-agent visualization generation approach. 
We use high-level visualization instructions that reflect realistic user intents, without explicitly specifying workflow construction details.




The baseline configuration adopts C5 from Experiment~\ref{sec:exp1}. 
We then apply the proposed multi-agent framework to generate workflows from the user query. 
As described in Section~\ref{sec:multistages}, the process is decomposed into three stages: intent planning (Stage~1), workflow template generation (Stage~2), and self-correction (Stage~3).

Due to ambiguity in the input instructions, the generated workflows may vary across runs. 
We therefore conduct three independent runs \textcolor{review}{to evaluate} the stability of the generated results.
For each run, the LLM first generates a workflow template, which is then manually refined into a runnable version using a dedicated visualization interface. 
We compare the generated workflows with the refined versions to quantify the amount of correction required, following the same metrics as the previous experiment. 
Detailed results and the visualization interface are provided in the supplementary materials.

\begin{figure}[htp]
    \centering \includegraphics[width=0.76\columnwidth]{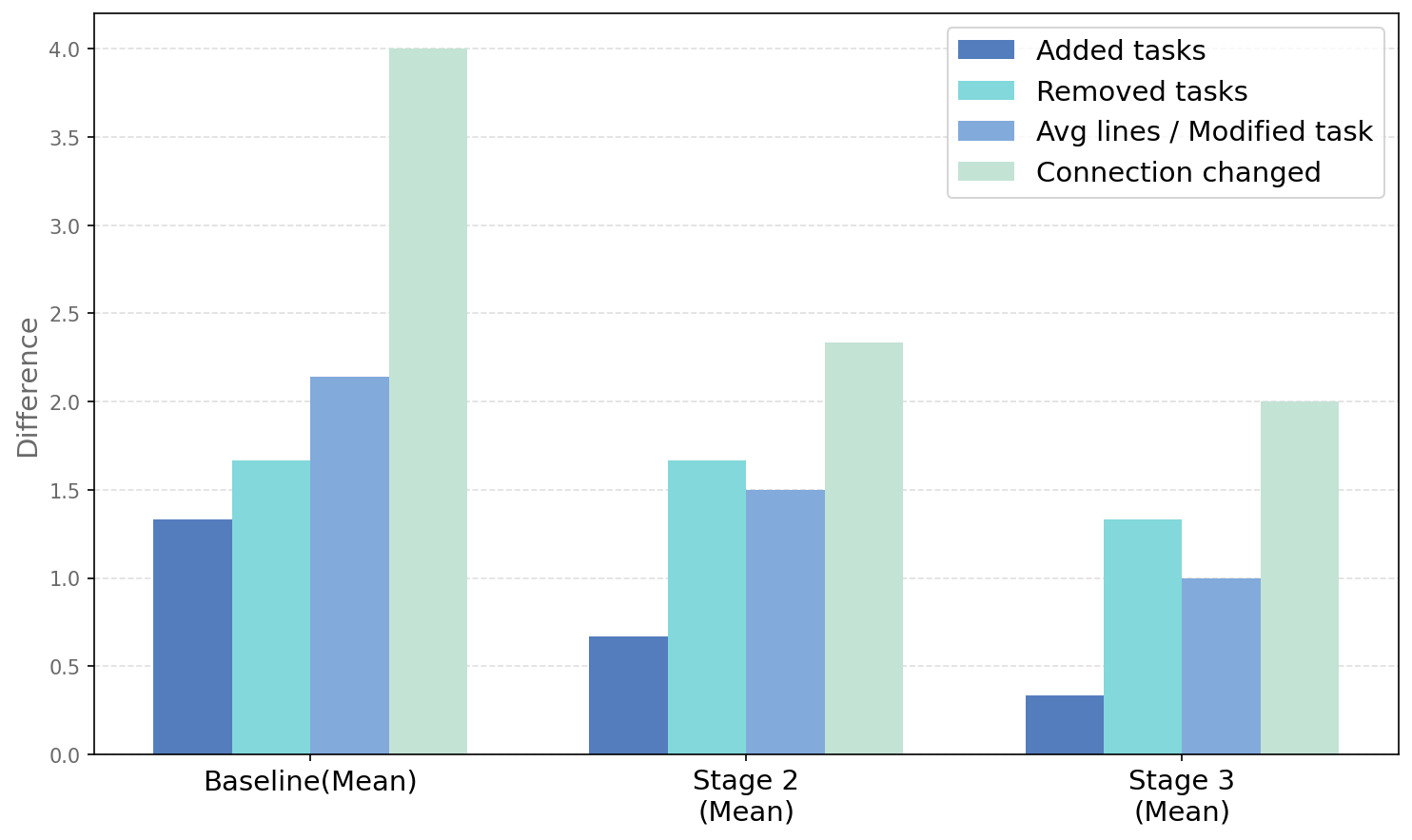}
    \caption{\textcolor{review}{Workflow template differences: baseline vs. multi-stage generation (lower bounds truncated at 0).}}
    \label{fig:exp2_diff}
    \vspace{-10pt}
\end{figure}

Figure~\ref{fig:exp2_diff} shows detailed differences between generated and corrected workflows \textcolor{review}{for the baseline, Stage~2, and Stage~3 averaged over three runs. Detailed experiment number are listed in the supplementary materials.} 
The results indicate that both Stage~2 and Stage~3 outperform the baseline across all runs, and Stage~3 achieves the best performance. 
Test 3 in Stage~3 even produces a near-correct workflow, with no added or removed nodes, and only two camera parameters are adjusted.
This improvement is attributed to the error-aware refinement mechanism, which captures common failure patterns and reduces repeated mistakes. 
Furthermore, the multi-agent decomposition reduces the need for extensive operator descriptions in the input, resulting in more compact prompts and lower context overhead. 
For example, the total token usage decreases from 23,500 to 14,000 approximately when using the multi-stage approach to generate a workflow template.
















\section{Conclusion and Future work}

This paper presents an intelligent workflow generation approach for turbomachinery visualization that integrates structured representations, domain-aware context knowledge, and multi-stage agent reasoning. The proposed method translates natural language intents into intermediate representations and supports interactive refinement. Results show improved reliability and reduced manual effort. Future work will extend the framework to more turboachinery visualization scenarios, such as turbines and combustors, standardized workflow interfaces, and improved robustness through better context knowledge management and multi-pass verification.

\section*{Acknowledgment}
\textcolor{review}{This work was supported by the Beijing Natural Science Foundation (No.4254090) and the Youth Fund of Computer Network Information Center of Chinese Academy of Sciences (No.25YF07). The authors gratefully acknowledge the experts at the Digital Twin Research Center, Institute of Engineering Thermophysics, Chinese Academy of Sciences, and Zhonghao Wang (Institute of Engineering Thermophysics) for providing constructive suggestions and assistance for the user study.}


\bibliographystyle{abbrv-doi}
\bibliography{main}

\end{document}